\documentclass{iauc}
\usepackage{graphicx}
\newcommand\Msun {M_{\odot}}

\title[Local Group Dwarf Spheroidals]{The Chemical Composition of Local 
Group Dwarf Spheroidals}

\author[Tolstoy]{Eline Tolstoy}

\affiliation{Kapteyn Institute, University of Groningen,
Postbus 800, 9700AV Groningen, the Netherlands\break 
\\}

\pubyear{2005}
\volume{198} 
\pagerange{1--8}
\date{1st June 2005}
\jname{Near-field Cosmology with Dwarf Elliptical Galaxies}
\editors{H. Jerjen \& B. Binggeli eds.}
\begin{document}

\maketitle

\begin{abstract}

I will review the progress of VLT 
spectroscopy of large numbers of individual stars in nearby dwarf
spheroidal galaxies. This spectroscopy has allowed us to obtain
detailed insights into the chemical and dynamical properties of the
resolved stellar population in these nearby systems.

\end{abstract}

\firstsection % if your document starts with a section,
\section{Introduction}

The dwarf galaxies being considered here are the lowest luminosity
(and mass) galaxies that been have found.  It is perhaps not a surprise that
there is a lower limit in 
mass for an object to be able to form more than one
generation of stars which will be related to
the limit below which one supernova will
completely destroy a galaxy. Numerical and analytic models tell us
that the limit must be around a few $\times 10^6 \Msun$ 
(e.g., Ferrara \& Tolstoy 2000).  
Globular clusters, for example, typically have much
lower masses than this limit.

It is among dwarf galaxies, irregular (dI) and spheroidal (dSph),
that we find the lowest mass objects. For example the Sculptor dSph galaxy
which apparently has a present day mass of 6$\times 10^6 \Msun$, or
the dI galaxy DDO~210 which has a present day mass of 5.4$\times 10^6
\Msun$ (Mateo 1998).  The only difference between the lowest mass
dSphs and dIs seems to be the presence of gas and current star
formation in dIs.  It has already been noted that dSphs predominantly
lie close to our galaxy ($< 250$kpc away), and dIs predominantly
further away ($>$400 kpc away). This suggests that the proximity
of dSph to our Galaxy played a role in the removal of gas from
these systems. Although the range of properties found in dSph and dI
galaxies does not allow a straight forward explanation; particularly
not the large variations in star formation histories and chemical
evolution paths that have now been observed in these systems.

%Figure 1
\begin{figure}[!ht]
{\centering \scalebox{0.6}{
\includegraphics[angle=270,scale=.75]{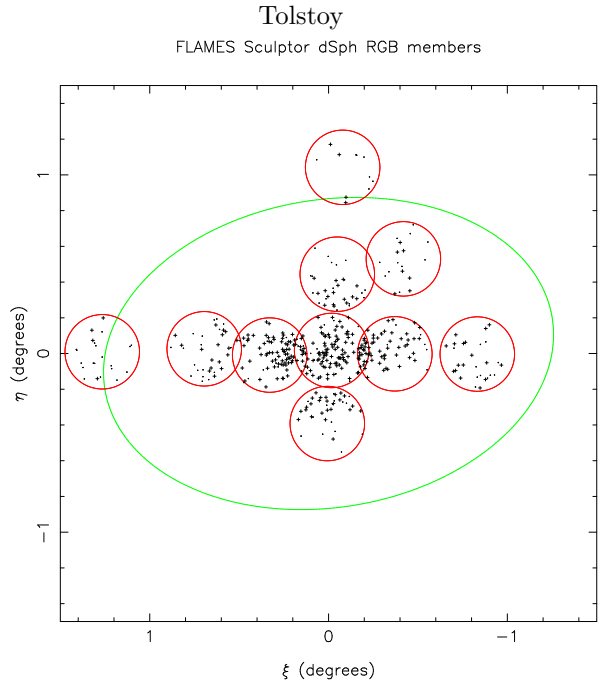}} \par}
\caption{
The distribution of candidate members RGB stars (shown as small
crosses) and velocity non-members (dots) in the Sculptor dSph
galaxy chosen from velocities measured in VLT/FLAMES
spectra.  The ellipse is the tidal radius of Sculptor, determined by
Irwin \& Hatzidimitriou (1995). The circles represent the 10
individual FLAMES pointings made by DART in this galaxy.
}
\end{figure}

Here I am going to discuss some recent observations of individual
stars in nearby dSph galaxies, predominantly from the large survey
being undertaken by DART (Dwarf Abundances and Radial-velocities Team)
to measure abundances and velocities for several hundred individual
stars in a sample of three nearby dSph galaxies: Sculptor,
Fornax and Sextans.  We have used the VLT/FLAMES facility to obtain
more than 100 spectra per 25arcmin diameter field of view across
several positions in each galaxy out to the tidal radius (see Figure 1).
We have used the low resolution setting to obtain Ca~II triplet
metallicity estimates as well as accurate radial velocity measurements
(Tolstoy et al. 2004; Battaglia et al., in prep) over a large area in
each galaxy, and we have also used the high resolution settings to
concentrate on about 100 stars in the central field of each galaxy -
where we are able to obtain accurate abundances for a range of
interesting elements such as Mg, Ca, O, Ti, Na, Eu to name a few (Hill
et al., in prep; Letarte et al., in prep).

\section{How do we understand the evolution of galaxies}

%Figure 2
\begin{figure}[!ht]
{\centering \scalebox{0.6}{
\includegraphics[scale=.8]{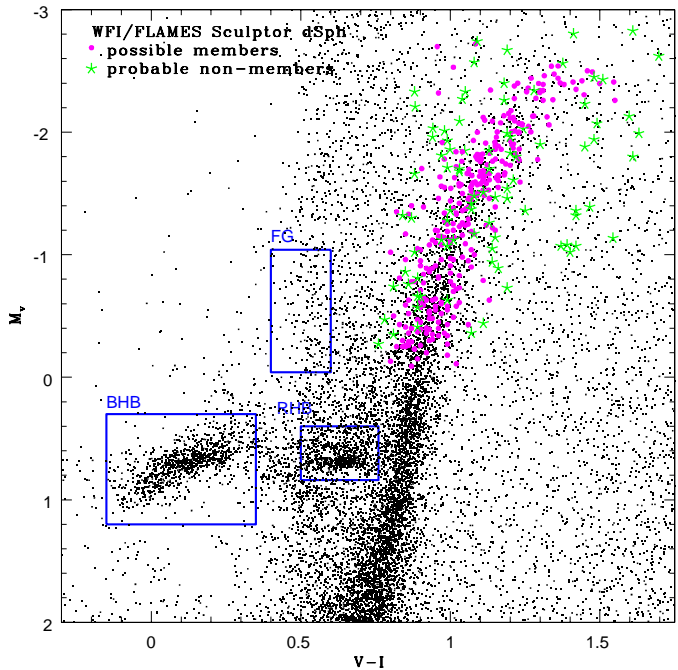}} \par}
\caption{
The Colour-Magnitude Diagram for the WFI coverage of Scl shown in
Figure~1.  The circle and star symbols are the Red Giant Branch stars
which were observed with VLT/FLAMES and for which we have accurate
v$_{hel}$ and [Fe/H] measurements (S/N$>$10).  The potential members
of Scl are shown as circles and non-members shown as stars.  Also
shown are the regions used to define the Blue Horizontal branch
(BHB), the red horizontal branch (RHB) and foreground
comparison (FG) populations.
}
\end{figure}

When we have resolved the individual stars in a galaxy and accurately
measured their colours and magnitudes down to the oldest main sequence
turn-offs (M$_V \sim +3.5$) in a Colour-Magnitude Diagram we can in
principle come to detailed conclusions about how the star formation
rate has varied with time all the way back to the first star formation
in the galaxy (e.g., Tolstoy \& Saha 1996; Gallart et al. 1999;
Hurley-Keller et al. 1999). This approach is the most accurate for
intermediate age populations, but for stars older than about 10~Gyr
the time resolution gets quite poor (and the stars are getting very
faint), and it becomes hard to distinguish a 12~Gyr old star from a
10~Gyr old star. Here it becomes useful to consider the Horizontal
Branch stars (M$_V \sim 0.$) which are the bright He-burning phase of
low mass stars $>$10~Gyr old. The ratio of red to blue horizontal
branch stars tells us about the age and metallicity variation 
(Lee et al. 2001) of the
oldest stars in the galaxy (e.g., for Sculptor dSph, see Figs 2 \& 3).

The Red Giant Branch (RGB, $-3 < M_V < 0.$) contains stars with ages
$>$1~Gyr old, back to the oldest stars in the galaxy.  In the majority
of RGB stars it is believed that the atmosphere of the star remains an
unpolluted sample of the interstellar medium out of which it was
formed. This means we have small pockets of interstellar medium of
different ages (enriched by different numbers of processes)
conveniently covering the nuclear burning core of stars. This hot
stellar core provides a useful bright background source to be absorbed
in the stellar atmosphere allowing very detailed studies of the
elemental abundances in these ancient gas samples.  Thus, a
spectroscopic analysis of the variation of the abundances of different
elements seen in absorption in atmospheres of different age stars
allows us to trace the detailed chemical enrichment history of a
galaxy with time.

%Figure 3
\begin{figure}[!ht]
{\centering \scalebox{0.8}{
\includegraphics[angle=270,scale=.75]{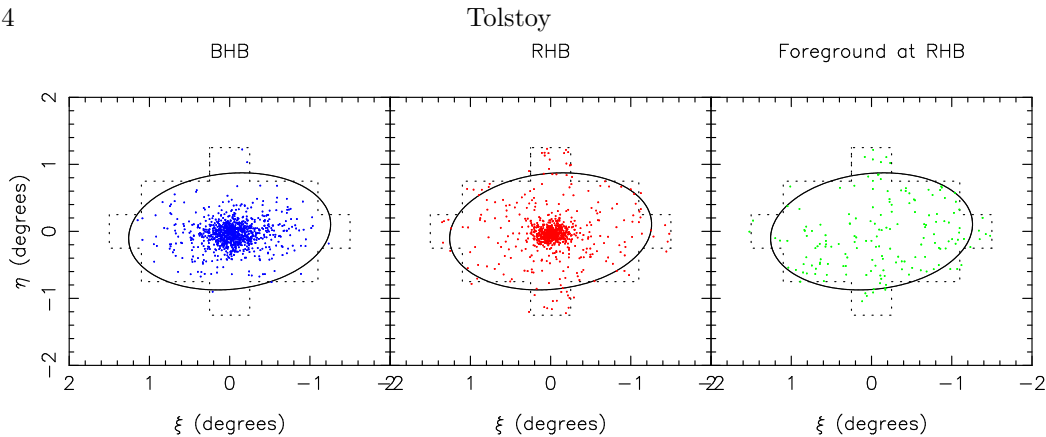}} \par}

\caption{
The distribution of Horizontal Branch stars from ESO/WFI imaging of
the Scl dSph.  The three panels show the different spatial
distributions of blue horizontal branch stars (BHB), and red
horizontal branch stars (RHB) as selected from the M$_{v}$, V$-$I
Colour-Magnitude Diagram (CMD) shown in Figure 2.  Also shown, to
illustrate the foreground contamination in the RHB distribution, are a
CMD-selected sample of foreground stars to match the RHB contamination
density (see Figure 2).  The ellipse is the tidal radius of Scl as
defined by Irwin \& Hatzidimitriou (1995).
}

\end{figure}

The age-metallicity degeneracy on the RGB (graphically illustrated by
Cole et al. 2005; their Figure 8) leads to serious problems using
photometric colour alone to disentangle age and metallicity.  This
uncertainty is the leading reason for quite different star formation
histories to be inferred from the same Colour-Magnitude Diagrams.  For
example, in the case of the Carina dSph three groups determined three
different star formation histories from two different data sets
(Hurley-Keller et al. 1998; Hernandez et al. 2000; Dolphin
2002), some of the (significant) differences can be traced to different
assumptions about the chemical evolution of Carina.  However if
similar chemical enrichment histories are assumed then the consistency
between models by different groups is significantly improved (e.g.,
Skillman et al. 2003).

The observed magnitude and colour (e.g., M$_V$, V$-$I) of a star
combined with a measured metallicity, [Fe/H], allows us to effectively
remove the age-metallicity degeneracy and determine the age of an RGB
star from an isochrone.  However the uncertainties in this age
increase with {\it decreasing} metallicity, and often the photometric
errors on the colour measurements are of the same order of magnitude
as the entire age range of the isochrones from the oldest (13~Gyr) to
the youngest (2~Gyr).  This is before even considering other effects
such as the variations in the Galactic reddening, $\alpha$-element
abundance, distance, metallicity measurement error or errors in the
interpolation of the models.

The interpretation of all these stellar measurements rely upon stellar
evolution models which are primarily based upon our understanding of
Galactic globular cluster stars and we must make the assumption that
this knowledge is generally applicable to stars in dwarf galaxies even
though the initial abundance ratios are different, and certainly the
stellar density is very different.

\section{High Resolution Spectroscopy: Detailed Abundance Analysis}

The best way to measure the chemical 
evolution of a galaxy
is to determine 
the relative abundances of different chemical
elements. This has been called chemical tagging by Freeman \&
Bland-Hawthorn (2002). Different elements are created in different
circumstances and if we are able to determine the abundance of elements known
to be created in a particular set of physical conditions, we can assess
the importance and frequency of these conditions during the history of
star formation in a galaxy.  This clearly can give insights into the
star formation rates at different times in a galaxy's history and the
environments in which most of the stars were formed.  

For example, the abundance of light elements (e.g., O, Na, Mg, Al) are
considered to be tracers of ``deep mixing'' patterns which
are found only in globular cluster environments, which gives
a limit to the number of dissolved globular clusters which can 
exist in a stellar population. These typical (Galactic) globular
cluster abundance patterns have also been recently found in the
globular clusters of Fornax dSph (Letarte et al. 2005), but not (so far)
in the field star populations (Shetrone et al. 2003).

%Figure 4
\begin{figure}[!ht]
{\centering \scalebox{0.8}{
\includegraphics[scale=.8]{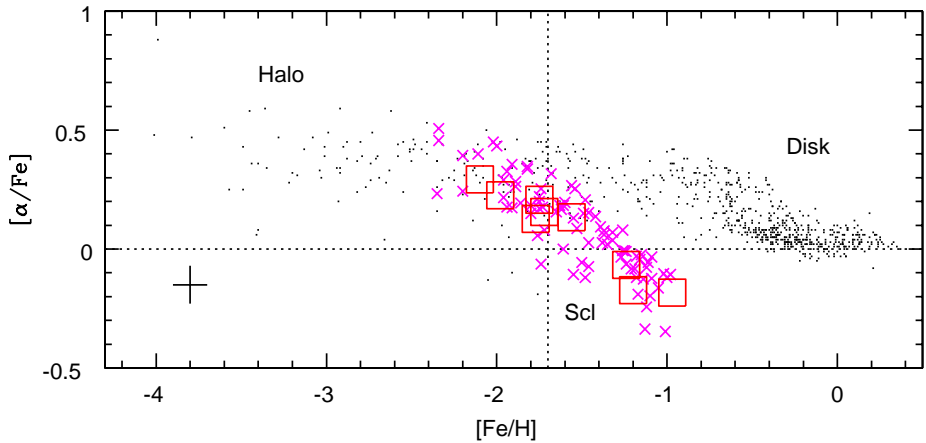}} \par}
\vskip -6.5cm
\caption{
The $\alpha$-abundance (average of Ca, Mg and Ti) for stars in our
Galaxy compared to those in Scl. The VLT/FLAMES high resolution
measurements of 92 members in the central field are shown as crosses
(from Hill et al., in prep).  The Galactic stars come from standard
literature sources (see Venn et al. 2004 for references). The 10 open squares
are UVES measurements of individual stars in Scl (Shetrone et al. 2003; 
Geisler et al. 2005).
}
\end{figure}

The creation of $\alpha$-elements (e.g., O, Mg, Si, Ca, Ti) occurs
predominately in Supernovae type II explosions, i.e. the explosion of
massive stars a few 10$^6 - 10^7$ yrs after their formation. The
abundance of the different $\alpha$-elements is quite sensitive to the
mass of the SNII progenitor (e.g., Woosley \& Weaver 1995) so the
$\alpha$/Fe ratio traces the mass function of the stars which
contributed to the creation of the $\alpha$ elements, and ratios of
different $\alpha$ elements themselves can put limits on the highest
mass star which has enriched a galaxy and also the typical mass range
(e.g., McWilliam 1997).

Heavy Elements (Z $>$ 30) are a mix of r- and s- process
elements. That is to say elements which were produced by rapid or slow
neutron capture which tells us about the environment in which
enrichment occurred.  Rapid capture (r-process) is assumed to occur in
high energy circumstances, such as supernovae explosions. For example Eu is
considered to be an element produced almost exclusively by the
r-process. The slow capture (s-process) is thought to be created by
more quiescent processes such as the stellar winds common in AGB type
stars of intermediate age (and mass). Typical elements which are
thought to be created by the s-process are Ba and La. The ratio between
r- and s- process element abundances gives an indication of the
relative importance of these different enrichment processes during the
history of star formation in a galaxy. Dwarf spheroidal galaxies are
generally found to have strong s-process enhancement. This may
indicate that supernovae products are typically lost to a shallow
potential well, or that the slow star formation rate means that
massive stars are not very common (Tolstoy et al. 2003; Venn et al. 2004).

With an 8m class telescope it is possible to measure the abundance of
these elements in dSph galaxies in a few hours of
integration time with a high-resolution high-throughput spectrograph,
e.g.  KECK/HIRES (Shetrone et al. 2001); VLT/UVES (Tolstoy et
al. 2003; Shetrone et al. 2003; Geisler et al. 2005; Letarte et
al. 2005, in prep).  But it has been quite slow and laborious to obtain a
statistically meaningful sample of stars in dSph galaxies which have
fairly complicated star formation histories. This complex star forming
history means that a sizeable sample of stars of different ages is
required to trace the full range enrichment processes.  The real 
break-through comes with VLT/FLAMES (and other fibre systems, such as MIKE
on Magellan). In one shot spectra can be obtained for over 100 stars
in one 25arcmin field of view. The sensitivity and the field of view
are a very nice match to the Galactic dSph galaxies.

The results published to date still lack good statistics to make a
detailed comparison between our Milky Way and individual dSph, although
attempts have been made (e.g, Tolstoy et al. 2003; Venn et al. 2004).
From these results, and the early results from our FLAMES surveys
(e.g., Hill et al, in prep; Battaglia et al., in prep) a match is not
found between the detailed abundance properties of stellar populations
found in dSph galaxies and any (significant) component of our
Galaxy. There remains a discrepancy between the metallicity ([Fe/H])
distribution of the stars randomly selected in a dSph (e.g., Scl:
Tolstoy et al. 2004; also true for Fornax and Sextans, Battaglia et
al. in prep) and samples of stars in our Milky Way. This is also true
when more detailed comparisons are made, for example considering how
[$\alpha$/Fe] varies with [Fe/H] (Hill et al. 2004, in prep; see
Figure 4). 

%Figure 5
\begin{figure}[!ht]
{\centering \scalebox{0.8}{
\includegraphics[scale=.8]{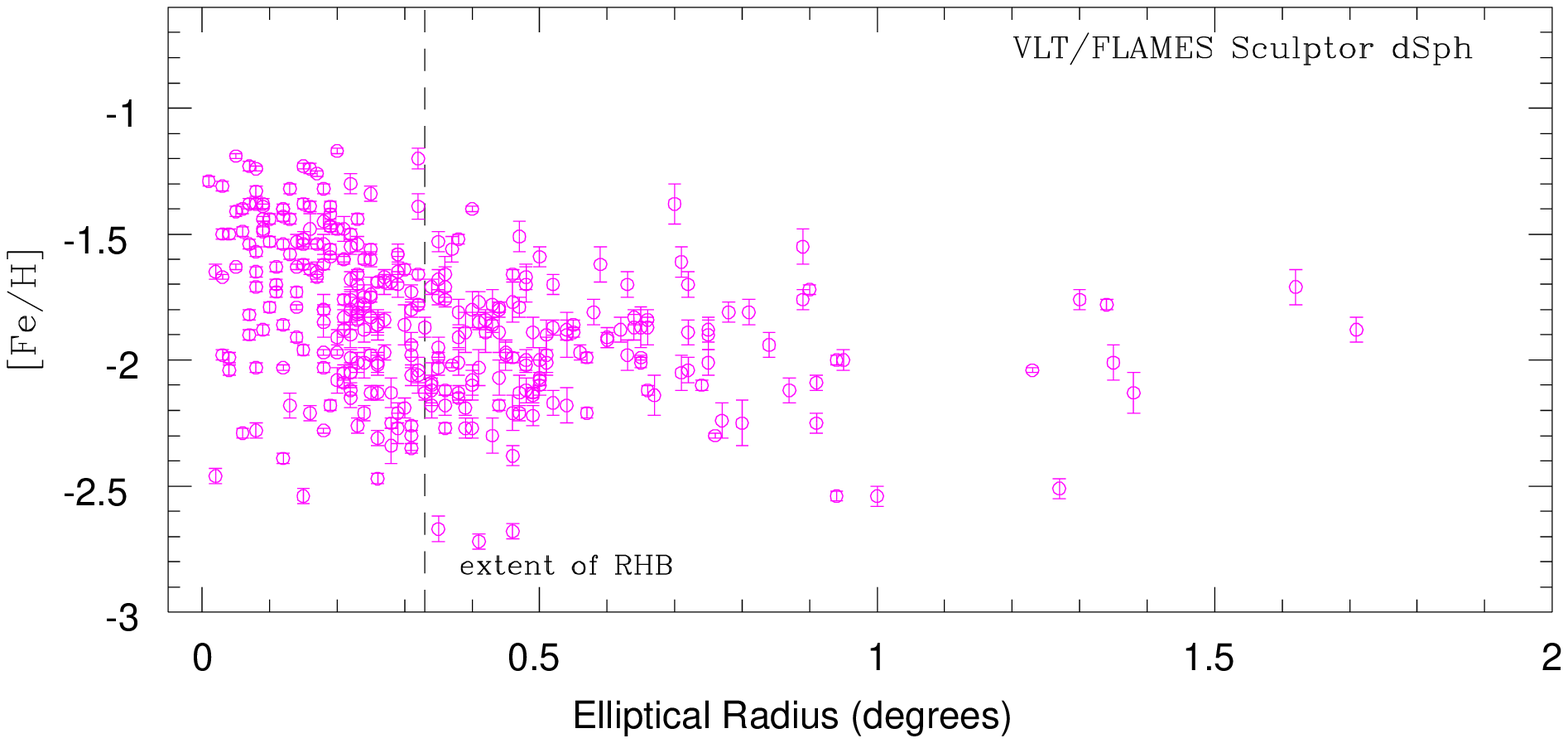}} \par}
\vskip -13.cm
{\centering \scalebox{0.8}{
\includegraphics[scale=.8]{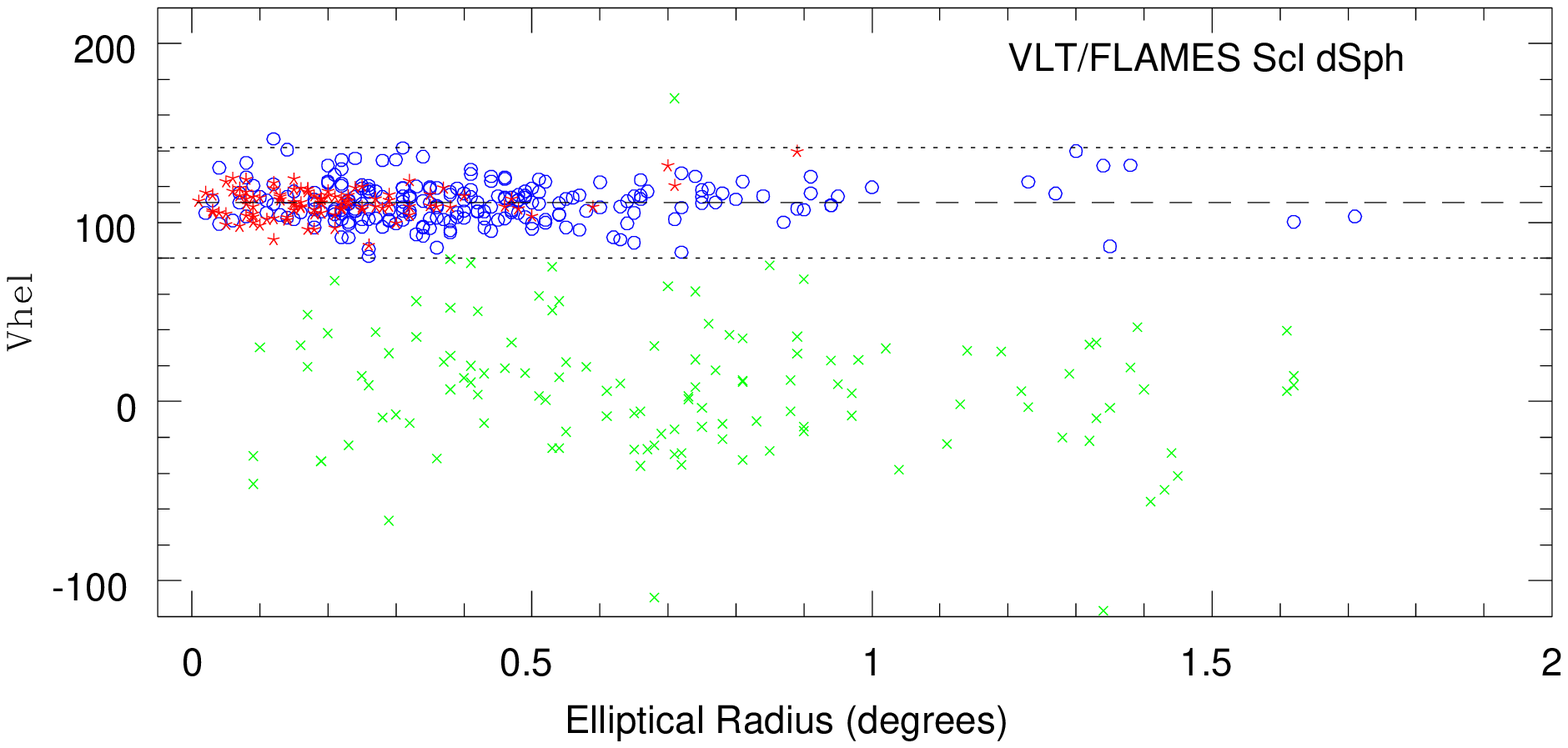}} \par}
\vskip -.5cm
\caption{
The upper panel shows the VLT/FLAMES spectroscopic measurements of
[Fe/H] for the 321 potential RGB velocity members of Scl (and
S/N$>$10).  We see a clear trend of metallicity with radius.  The
lower panel shows v$_{hel}$ as a function of elliptical radius for all
stars satisfying S/N $>$ 10.  Likely Scl members are clearly seen
clustered around the systemic velocity of 110 km/s. The 321 stars
which are potential members are plotted as stars ([Fe/H] $>$ -1.7) and
circles ([Fe/H] $<$ -1.7), while the 128 crosses are assumed to be
non-members.
}
\end{figure}

\section{Low Resolution Spectroscopy: Metallicities}

The ideal is to be able to obtain high resolution spectra for
individual stars in nearby dSph over a large wavelength range, and
make a detailed analysis of a range of different elements. However, 
this is quite time consuming both in telescope time (at the distance
of dSph) and in analysis.  One of the most simple ways to get an
estimate of the metallicity of RGB stars is with the Ca~II triplet
(e.g., Cole et al. 2004). This is a basic metallicity indicator
requiring only low or intermediate spectral resolution, based on three
lines around 8500\AA which have been empirically calibrated from
observations of stars in globular clusters stars with high resolution
abundances (e.g., Armandroff \& Da~Costa 1991).

With the Ca~II triplet we obtain an overview of the global metallicity
range of the RGB stars in a dwarf galaxy. In the case of the DART
project these measurements are complementary to the high
resolution observations made in the centre of each dSph. In the low
resolution larger area survey we can assess how representative our
detailed study is of the stellar population of the whole galaxy.

%Figure 6
\begin{figure}[!ht]
{\centering \scalebox{0.85}{
\includegraphics[angle=270,scale=.65]{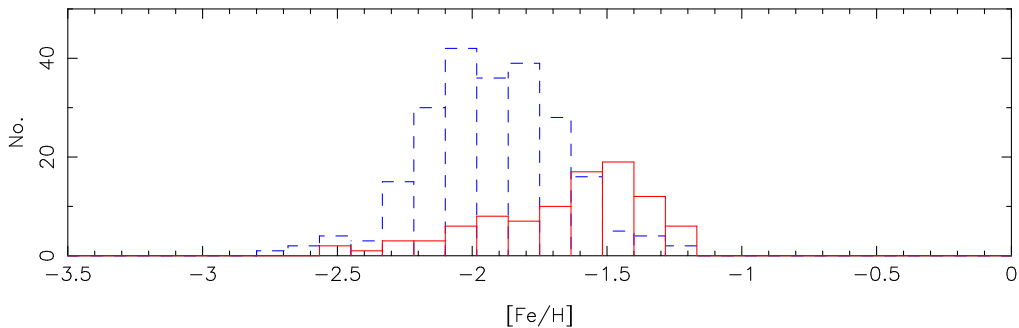}} \par}
%\vskip -.5cm
\caption{
Here is the histogram distribution of [Fe/H] for Scl dSph: 
the 94 stars 
within the central, r$<$0.2 degree region (solid line); and the 227 
stars beyond r$>$0.2 degrees (dashed line). 
}
\end{figure}

Our first VLT/FLAMES results, based upon Ca~II triplet, clearly show
that Scl dSph contains two distinct stellar components with different
spatial, kinematic and abundance properties (Tolstoy et al. 2004, see
Figure 5).  There appears to be a metal-rich, $-0.9 >$ [Fe/H] $>
-1.7$, and a metal-poor, $-1.7 >$ [Fe/H] $> -2.8$ component. The
metal-rich component is more centrally concentrated than the metal
poor, and on average appears to have a lower velocity dispersion,
$\sigma_{metal-rich} = 7 \pm 1$ km/s, whereas $\sigma_{metal-poor} =
11 \pm 1$ km/s.

The full abundance analysis of the FLAMES HR data (Hill et al. in
prep) will provide more details of the chemical enrichment history of
Scl. This will enable us to distinguish between two episodes of star
formation or more continuous star formation, manifested as a gradient
in velocity dispersion and metallicity from the centre of the galaxy.

It is clear from the histogram of [Fe/H] measurements in Scl dSph (see
Figure 6) that this distribution lacks a low metallicity tail, in fact
the lowest metallicity star in our sample of more than 300 stars is
[Fe/H] $ = -2.7$. We find a similar lack of low metallicity stars in
Fornax and Sextans.  Although it is difficult to make an accurate
comparison with Galactic samples, where the completeness of different
samples is hard to quantify, there appears to be a significantly
different distribution between dSph and the (metal-poor) halo of the
Milky Way.

\section{Conclusions}

There are indications that the presence of two populations is a common
feature of dSph galaxies.  Our preliminary analysis of Horizontal
Branch stars, v$_{hel}$ and [Fe/H] measurements in the other galaxies
in our sample (Fornax and Sextans dSph; Battaglia et al., in prep)
also show similar characteristics to Scl, especially in the most metal
poor component. Pure radial velocity studies (Wilkinson et al. 2004;
Kleyna et al. 2004) have also considered the possibility that
kinematically distinct components exist in Ursa Minor, Draco and
Sextans dSph galaxies.

What mechanism could create two ancient stellar components in a small
dwarf spheroidal galaxy?  A simple possibility is that the formation
of these dSph galaxies began with an initial burst of star formation,
resulting in a stellar population with a mean [Fe/H] $\leq - 2$.
Subsequent supernovae explosions from this initial episode could have
been sufficient to cause gas (and metal) loss such that star formation
was inhibited until the remaining gas could sink deeper into the
centre (e.g., Mori et al. 2002).  Thus the subsequent generation(s) of
stars would inhabit a region closer to the centre of the galaxy, and
have a higher average metallicity and different kinematics.  Another
possible cause is external influences, such as minor mergers,
accretion of additional gas or the kinematic stirring by 
our Galaxy.  It might also be that
events surrounding the epoch of reionisation influenced the evolution
of these small galaxies (e.g., Skillman et al. 2003) and resulted in
the stripping or photo-evaporation of the outer layers of gas in the
dSph, meaning that subsequent more metal enhanced star formation
occurred only in the central regions.

\begin{acknowledgments}
I am grateful for support from a fellowship of the Royal Netherlands
Academy of Arts and Sciences, and the exceptional collaborators that
make up DART, especially Mike Irwin, Giuseppina Battaglia, 
Amina Helmi, Andrew Cole, Vanessa Hill, Kim Venn,
Bruno Letarte, Pascale Jablonka, Matthew Shetrone \& Andreas Kaufer.
\end{acknowledgments}


\begin{thebibliography}{}
\addcontentsline{toc}{section}{References}

\bibitem[]{adc91} Armandroff \& Da Costa 1991 AJ, 101, 1329

\bibitem[]{cole04} Cole A.A., Smecker-Hane T.A., Tolstoy E. et al. 2004 MNRAS, 347, 367

\bibitem[]{cole05} Cole A.A., Tolstoy E., Gallagher J.S. \& Smecker-Hane T.A. 2005 AJ, 129, 1465

\bibitem[]{dolp02} Dolphin A. 2002 MNRAS, 332, 91

\bibitem[]{ft00} Ferrara A. \& Tolstoy E. 2000 MNRAS, 313, 291 

\bibitem[]{freebh02} Freeman K. \& Bland-Hawthorn J. 2002 ARA\&A, 40, 487

\bibitem[]{gall99} Gallart C., Freedman W.L., Aparicio A., Bertelli G. \& Chiosi C. 1999 AJ, 118, 2245

\bibitem[]{geis05} Geisler, D., Smith, V., Wallerstein, G., Gonzalez, G. \& 
Charbonnel, C. 2005 AJ, 129, 1428

\bibitem[]{her00} Hernandez X., Gilmore, G. \& Valls-Gabaud, D. 2000 MNRAS, 317, 8831

\bibitem[]{hk98} Hurley-Keller D., Mateo M., \& Nemec J. 1998 AJ, 115, 1840

\bibitem[]{hk99} Hurley-Keller D., Mateo M., \& Grebel E.K. 1999 ApJL, 523, 25

\bibitem[]{ih95} Irwin M. \& Hatzidimitriou D. 1995 MNRAS, 277, 1354

\bibitem[]{k04} Kleyna J.T., Wilkinson M.I., Evans N.W. \& Gilmore G. 
2004 MNRAS, 354, L66

\bibitem[]{lee01} Lee Y-W. et al. 2001 {\it Astrophysical Ages \& Time Scales}, eds. T. von Hippel et al.

%\bibitem[]{Maj99} Majewski, S.R., Siegel, M.H., Patterson, R.J. \& Rood, R. 1999 ApJL, 520, 33

\bibitem[]{mateo98} Mateo M. 1998 ARA\&A, 36, 435

\bibitem[]{mcw97} McWilliam A. 1997 ARA\&A 35, 503

\bibitem[]{mori02} Mori M., Ferrara A. \& Madau P. 2002 ApJ, 571, 40

\bibitem[]{shet01} Shetrone M.D., Cote P. \& Sargent W.L.W. 2001, ApJ, 548, 592

\bibitem[]{shet03} Shetrone M.D., Venn K.A., Tolstoy E., Primas F., Hill V. \& Kaufer A. 2003 AJ, 125, 684

\bibitem[]{skill03} Skillman E.D., Tolstoy E., Cole A.A., Dolphin A.E. et al. 2003 ApJ, 596, 253

\bibitem[]{ts96} Tolstoy E. \& Saha A. 1996 ApJ, 462, 672

\bibitem[]{et03} Tolstoy E., Venn K.A., Shetrone M.D.,  Primas F., Hill V. et al. 2003 AJ, 125, 707

\bibitem[]{et04} Tolstoy E., Irwin M.J., Helmi A., Battaglia G., Jablonka P., Hill V. et al. 2004 ApJL, 617, 119

\bibitem[]{venn04} Venn K., Irwin M., Shetrone M.D., Tout C.A., Hill V. \& Tolstoy E. 2004 AJ, 128, 1177

\bibitem[]{wilk04} Wilkinson M., Kleyna J.T., Evans N.W., Gilmore G. et al. 2004 ApJL, 611, 21

\end{thebibliography}
\end{document}